\documentstyle[epsf]{article}
\newcommand{\bee}{\begin{equation}}
\newcommand{\ee}{\end{equation}}
\newcommand{\beea}{\begin{eqnarray}}
\newcommand{\eea}{\end{eqnarray}}

\begin{document}
\thispagestyle{empty}
\parskip=12pt
\raggedbottom

\def\mytoday#1{{ } \ifcase\month \or
 January\or February\or March\or April\or May\or June\or
 July\or August\or September\or October\or November\or December\fi
 \space \number\year}
\noindent
\hspace*{9cm} COLO-HEP-383\\
\vspace*{1cm}
\begin{center}
{\LARGE  Topological Structure in the $SU(2)$ Vacuum}
\footnote{Work supported in part by
NSF Grant PHY-9023257 
and U.~S. Department of Energy grant DE--FG02--92ER--40672}

\vspace{1cm}

Thomas DeGrand,
Anna Hasenfratz,  and Tam\'as G.\ Kov\'acs\\
Department of Physics \\
 University of Colorado,
Boulder CO 80309-390

\vspace{1cm}

\mytoday \\ \vspace*{1cm}

\nopagebreak[4]

\begin{abstract}
We study the topological content of the vacuum of
   $SU(2)$ pure gauge theory using lattice simulations. We
use a smoothing process based on the renormalization group equation
which removes short distance fluctuations
but preserves long distance structure.  
The action of the smoothed configurations
 is dominated by instantons, but they still
show an area law for Wilson loops with a string tension equal to the
string tension on the original configurations.
Yet it appears that instantons are not directly responsible for confinement.
The average radius of an instanton
is about 0.2 fm, at a density of about 2 fm${}^{-4}$.
This is a much smaller average size than other lattice studies have indicated.
The instantons appear not to be randomly distributed in space, but
are clustered.
\end{abstract}

\end{center}
\eject


\section{Introduction}

An important ingredient in the physics of the strong interaction  
is the influence of topology on dynamics.  
Based on phenomenological models, it has been argued that instantons 
are largely responsible for the low energy hadron and glueball spectrum 
\cite{Diakonov,Shuryak_long}. Instanton liquid
models attempt to reproduce the topological content of the QCD 
vacuum and conclude that hadronic correlators in the instanton liquid 
show all the important properties of the corresponding full
QCD correlators. These models appear to capture the essence of
the QCD vacuum, but their derivations involve a number of uncontrolled 
approximations and phenomenological parameters.

Another example of this connection
is the $U(1)$ problem of QCD, whose resolution may involve instanton 
effects \cite{U1}.
In the large-$N_c$ limit the mass of the $\eta'$ is related to the
topological susceptibility  $\chi_t$ through
the Witten-Veneziano formula\cite{WV}
\bee
m^2_{\eta'} + m^2_{\eta} - 2m^2_K = 2 N_f \chi_t/f_\pi^2.
\label{WZF}
\ee
While the precise validity of this formula is unclear to us, it 
indicates the dynamical role of instantons in QCD.

Lattice methods are the only ones we presently have,
which might address this connection.
Individual instantons can be identified only on smooth enough
configurations therefore we perform
a smoothing process on the gauge fields which substantially reduces short
range fluctuations, but, as we prove and show explicitly, preserves
long distance physics. 
The smoothed configurations show an area law for Wilson loops
with a string tension equal to the string tension on the original
configurations.
We find that the action of these smoothed configurations
is dominated by instantons,
though tests we perform show that instantons are not directly
responsible for confinement.

Our goal in this paper is to study the topological 
structure of the pure SU(2) gauge vacuum. Our
method allows us to identify and study the properties of 
individual instantons on the lattice whose radius is as small 
as one lattice spacing. 
We find that the vacuum is filled with instantons whose average radius
is about 0.2 fm, at a density of about 2 fm${}^{-4}$.
This is a much smaller average size than other lattice studies
 have indicated \cite{MI_SPEN,Forcrand}. The instantons appear not to 
be randomly distributed in space, but are clustered.

\subsection {Topology on the lattice}

Lattice studies of topology suffer from the presence of lattice
artifacts. They can arise both from the form of the lattice action
and from the choice of lattice operator used to define and measure
topological charge. A lattice action is, in general, not scale invariant, 
i.e. the action of a smooth continuum instanton can depend on its size.
Difficulties also arise because the 
topological charge is not conserved on the lattice.
When the size of an instanton becomes small compared to the lattice
spacing, the instanton charge operator will miss it, 
it "falls through" the lattice.
If the size scale of physically relevant instantons is smaller than
the size scale at which the lattice simulation supports them,
one's predictions will exhibit scaling violations.
 
Fixed point actions allow us, in principle,
to solve the first of these problems. 
Predictions of physical observables computed from a lattice
action and  lattice operators
which live on the renormalized trajectory (RT) of some
renormalization group transformation (RGT)
do not depend on the lattice spacing, 
they are free of lattice artifacts.
 A recent series of
papers \cite{HN,PAPER1,PAPER2,PAPER3} has shown
how to find a fixed point (FP) action for asymptotically free theories,
with explicit examples for spin and gauge models. FP actions share the
scaling properties of the RT (through   one-loop quantum corrections)
and as such may be taken as a first approximation to a RT.

FP actions  also offer a way to define a topological
charge operator on the lattice, through the RGT equation itself.
Suppose we have a configuration of lattice variables 
$\{V\}$. We will refer to this configuration as coarse. We want to determine 
the topological properties of $\{V\}$ by mapping it onto 
a fine configuration $\{U\}$ (defined on a 
lattice with half the lattice spacing or
with $2^4$ times the original number of lattice points),
which has the same topology as $\{V\}$ but is 
smooth enough that the standard (geometrical or
algebraic) definition of the charge operators work. 
Such a mapping is possible through 
the steepest descent  FP equation 
\bee
S^{FP}(V)=\min_{ \{U\} } \left( S^{FP}(U) +\kappa T(U,V)\right),  \label{STEEP}
\ee
where $T(U,V)$ is the blocking kernel and the parameter 
$\kappa$ determines the "stiffness" of the
transformation \cite{HN,PAPER1,PAPER2,PAPER3}.

The fine configuration has the following properties:

1) The blocking kernel $T(U,V)$ is positive-definite,
therefore the action on the fine lattice
is always equal to or smaller than the action on the coarse lattice,
$S^{FP}(U)\le S^{FP}(V)$.

2) The configuration $\{U\}$  blocks into  $\{V\}$
under an RGT step. This statement is correct statistically for 
arbitrary RGT and becomes exact if $\kappa=\infty$.
That justifies referring to the transformation Eqn. \ref{STEEP} as inverse blocking.

3) As a consequence of 2), the long distance, physically 
relevant properties of $\{U\}$ and
$\{V\}$ are identical. In particular,
if a correlation length can be defined on a set of coarse lattices,
then the correlation length on the fine lattices, constructed by inverse
 blocking, will be the same as on the coarse lattices, 
when measured in physical units.
In addition to the physical, long distance structure, 
the fine configuration  $\{U\}$ exhibits a
peculiar staggered short range structure. 
This does not change its physical properties but
requires special care in measuring the spectrum. 
We will discuss this point in detail in Section 2.4.

4) The topological properties  of $\{U\}$ and $\{V\}$
are identical, at least for instantons that are larger 
than some fixed radius, $\rho_c$.

Neither the blocking  nor the inverse blocking transformations can 
create new instantons. The
blocking averages over small distance fluctuations, 
and the inverse blocking finds the minimum of
the action constrained by the blocking kernel. 
None of them allows the short scale fluctuations
necessary for creating a small instanton.

A  coarse configuration that is topologically trivial
will inverse block into a configuration with lower action, and after
many levels of inverse blocking the action on the finest lattice will
go to zero.

Next consider a smooth continuum instanton configuration.
As it was argued in \cite{HN}, if $\{ V \}$ is a solution of 
the classical equations of motion, so that
$\delta S^{FP}(V)/\delta V =0$, then Eqn. \ref{STEEP} says that
$\delta S^{FP}(U)/\delta U =0$: the configuration on the fine lattice is also
a solution to the classical equations of motion,
$S^{FP}(V)=S^{FP}(U)=S_I$ i.e. $\{ U \}$ is also a continuum instanton
configuration.
(The opposite is not necessarily true, 
since small instantons with radius less
then $\rho_c$ of the fine configuration could disappear after an RGT.)
The same is true for continuum multi-instanton configurations.
(When we describe multi-instanton configurations, we will frequently refer 
to both instantons and anti-instanton as instantons.)

For configurations  consisting of instantons/anti-instanton plus fluctuations, 
a similar argument works. The inverse blocking will not create new instantons; 
rather, it will decrease the action by smoothing out fluctuations. 
Since the resulting fine configuration blocks back to the
original coarse configuration, it has to have all the topology of the 
coarse configuration.

The size of the instantons grows by a factor of two under the inverse
blocking but their orientation and location is unchanged. 
After repeated application
of the transformation, the action will approach $S_I$ times the
total number of topological objects on the lattice.
Thus we can  define the topological charge of a
configuration by first inverse blocking
it to a sufficiently smooth configuration and then measuring the charge
with any appropriate operator on  the fine lattice.

The preceding paragraphs involved questions of principle.  The extent
to which they can be achieved in practice depends on our ability
to construct a good approximate FP action.
In our earlier papers \cite{INSTANTON1,INSTANTON2}
on this subject we found an approximate FP action for SU(2) pure
gauge theory, subjected it to scaling tests, and computed the topological
susceptibility. We saw that it scaled (within the uncertainties of our
simulations) for lattice spacings less than about 0.2 fm, at a value
of about $\chi_t^{1/4}=235$ MeV.  We have also presented some preliminary
measurements of the temperature dependence of the susceptibility
\cite{LATCONFINST}.

In this paper we present a new FP action with greatly improved scale
invariance and study its topological properties.
In Section 2 we introduce the procedures necessary 
to carry out the calculations.  In Section 3 we describe the properties
of instantons. Our conclusions are given in Section 4.

\section{Methodology}

\subsection{Construction of an Approximate FP Action}

The continuum (tree level) action of QCD is scale invariant,
i.e. instantons of different size have the same action.
This scale invariance is broken by the lattice cutoff, and the
way this happens may be crucial to the continuum limit.
The cutoff mainly influences  small
instantons with sizes around the lattice spacing. If the
action of these instantons decreases with their size (as is the
case e.g.\ with the Wilson action) then it can
happen that in the continuum limit topological observables will
be dominated by small instantons near the cutoff and
scaling for these observables will be lost. This was
first pointed out in the context of 2d spin models in
\cite{LU_BREAK,GEO_O3}
and in gauge theories in
\cite{PANDT,BREAKDOWN}.
 For a recent discussion of some related
questions in Yang-Mills theories see \cite{Polonyi}.
 
Topological charge operators can only detect  instantons above
a certain size, the size cutoff usually being of the order of the lattice
spacing. 
If this cutoff is above the size of the overabundant small instantons,
then  they will not be observed in
topological measurements. That does not mean, however,
that they can be neglected. Their presence can
affect the scaling of spectral quantities and cause scaling violations.

A classically perfect action is  scale invariant  by construction. 
In Ref. \cite{INSTANTON1} we presented a many-parameter FP action
for SU(2) pure gauge theory
which satisfied  our requirement that $S/S_I\ge 1$ for all charged
configurations, and $S/S_I=1$ for all smooth configurations
with topological charge $Q=1$.  This action was too unwieldy to 
use in simulations. We used two loops, the plaquette and the link-6 
twisted loop $(x,y,z,-x,-y,-z)$ to
construct an approximate FP action. This approximate action reproduced
 the FP action value within a couple of percent for Monte Carlo
configurations generated at large correlation length but was not
scale invariant for smooth instantons. In fact, it was only slightly better
than the Wilson action.
However, it did satisfy $S/S_I > 6/11$, so that
the entropic bound against overproducing dislocations was not violated.
We argued that this softer constraint should be adequate for practical
calculations.  However, in his studies of the $CP^3$ model \cite{CP3},
Burkhalter saw scale violations from the use of an action similar to ours, 
and in principle this action should also show scale violations
at large lattice spacing as discussed above.
In this paper we present an approximate FP action where  
scale violations are greatly reduced. 

The two loops used in Ref. \cite{INSTANTON2} are not sufficient to create a scale
invariant action. In our calculation we considered all loops with length
less or equal than 8 which fit into a $2^4$ hypercube. Among these 28 loops we
found that one, the perimeter-eight $(x,y,-x,-y,z,t,-z,-t)$ loop,
 is essential for constructing a scale invariant action.

\begin{figure}[!htb]
\begin{center}
\vskip 10mm
\leavevmode
\epsfxsize=90mm
\epsfbox{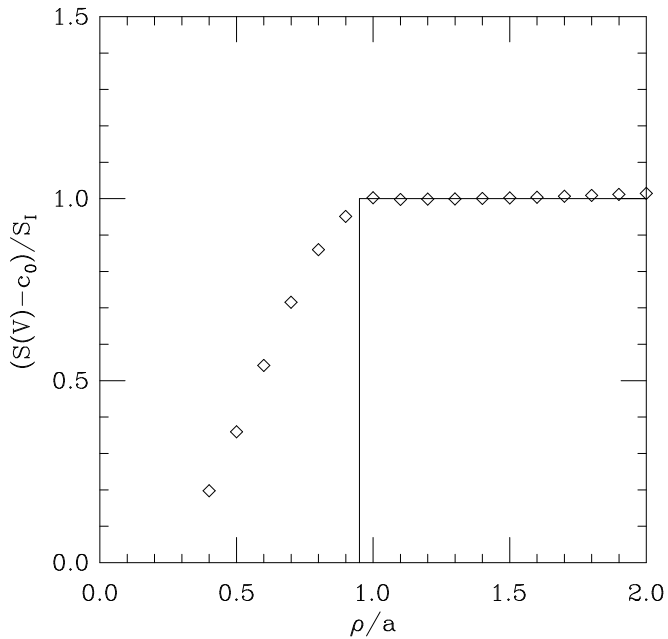}
\vskip 10mm
\end{center}
\caption{ Action vs. instanton radius for the SU(2) action of Table 1.}
\label{fig:newinstsu2}
\end{figure}

\begin{figure}[!htb]
\begin{center}
\vskip 10mm
\leavevmode
\epsfxsize=90mm
\epsfbox{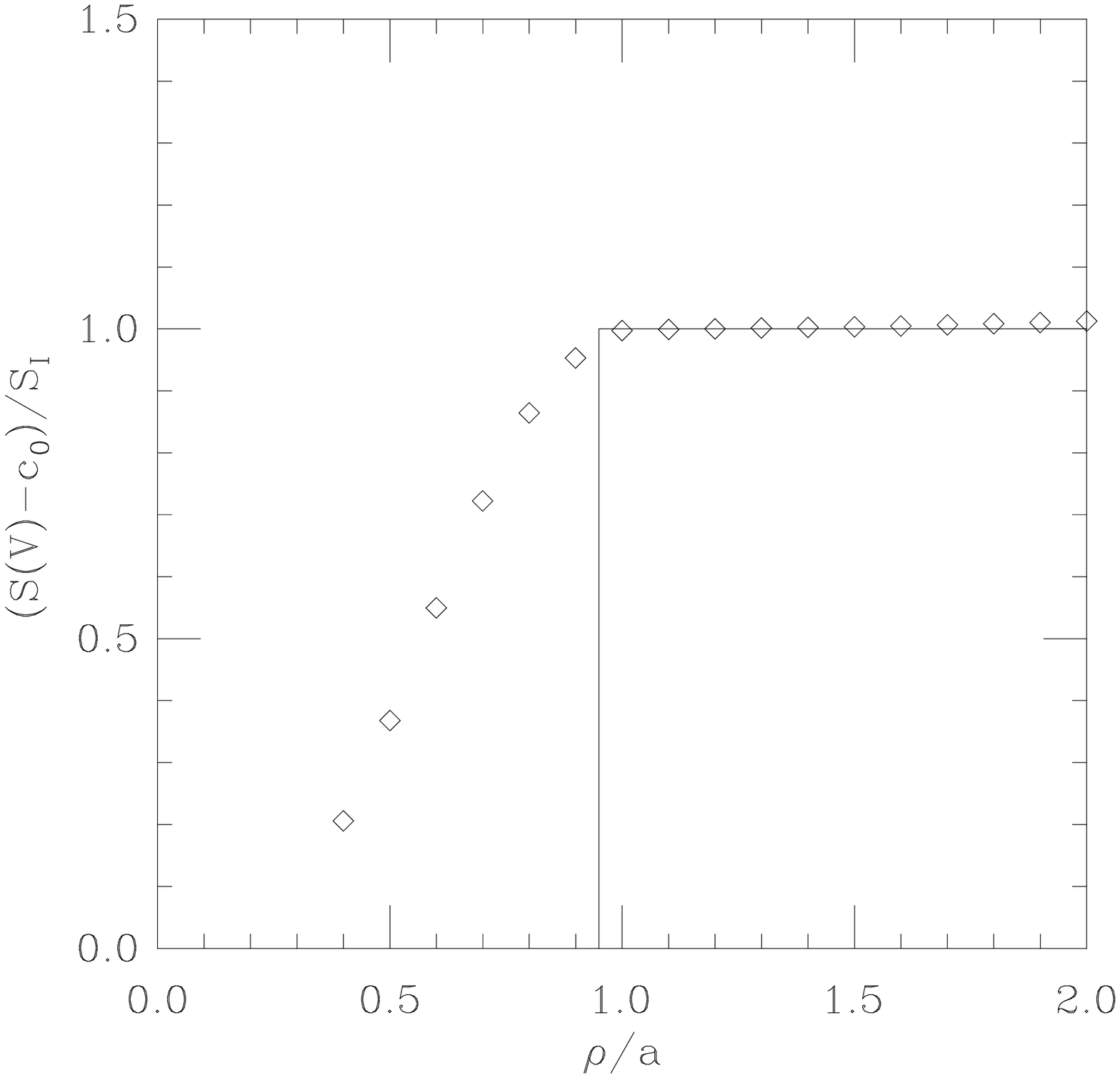}
\vskip 10mm
\end{center}
\caption{ Action vs. instanton radius for the SU(3) action of Table 2.}
\label{fig:newinstsu3}
\end{figure}

 Our new approximate FP action consists of four powers of
three loops, the plaquette,  the perimeter-six loop $(x,y,z,-x,-y,-z)$,
and the perimeter-eight loop $(x,y,-x,-y,z,t,-z,-t)$
\bee
S(V) = c_0+{1 \over N_c} \sum_{\it C} ( c_1({\it C})(N_c-{\rm Tr}(V_{\it C})) +
                       c_2({\it C})(N_c-{\rm Tr}(V_{\it C}))^2+ ...
\label{ACTIONPAR}
\ee
with coefficients tabulated in Table \ref{tab:eightparsu2}.
It is designed to fit ``typical'' gauge field configurations
in the range $aT_c= 1/3$ to 1/6, and, in addition, to preserve the 
scale invariance of pure isolated instanton configurations.

The constant term $c_0$ in the action is an unusual term, but it
 is necessary to fit
the FP equation Eqn. \ref{STEEP} in the small correlation length range
with a few parameters. 
If we were to attempt to find a parameterization of the FP
action valid for larger correlation length, the coefficient
 $c_0$ would change and eventually would
approach zero as $\beta \to \infty$. The $c_1-c_3$ coefficients in Eqn.
\ref{ACTIONPAR} are normalized such that $S-c_0$ has the correct continuum limit
and in a MC simulation $c_0$ is clearly irrelevant.
 To produce instantons with the
correct weight the action should give $S_I$ above the vacuum for smooth single
instanton configurations.
Fig. \ref{fig:newinstsu2} shows $(S(V)-c_0)/S_I$ vs. instanton radius
for a set of smooth instanton configurations. The action indeed scales within a
few percent even for small radius instantons.

We have not subjected the SU(2) action to a detailed scaling test
(test of  the variation of a dimensionless ratio of dimensionful observables
with lattice spacing) 
and we do not know if it has better scaling properties
than the action used in Ref. \cite{INSTANTON2} for spectral quantities.
However if
instantons are as important as the instanton liquid models predict, one
would expect improved scaling behavior at least for the glueballs.

\begin{table*}[hbt]
\caption{Couplings of the few-parameter FP action for SU(2) pure gauge theory.}
\label{tab:eightparsu2}
\begin{tabular*}{\textwidth}{@{}l@{\extracolsep{\fill}}lcccccc}
\hline
operator &  $c_1$ & $c_2$ & $c_3$ & $c_4$ & $c_0=-0.683$ \\
 \hline
$c_{plaq}$ &  .9225  &  .1804      & .0691 & -.0194   \\
$c_{6-link}$ &  -.1843  &  .1408 &  -.0264 & .0047        \\
$c_{8-link}$ &  .0485  &  -.0238 &  .0090 & .0026        \\
\hline
\end{tabular*}
\end{table*}

\begin{table*}[hbt]
\caption{Couplings of the few-parameter FP action for SU(3) pure gauge theory.}
\label{tab:eightparsu3}
\begin{tabular*}{\textwidth}{@{}l@{\extracolsep{\fill}}lcccccc}
\hline
operator &  $c_1$ & $c_2$ & $c_3$ & $c_4$  & $c_0=-1.669$ \\
 \hline
$c_{plaq}$ & 2.3350  &  -.4767      & .2465 & -.0207   \\
$c_{6-link}$ &  -.2391  &  .0565 &  .0305 & -.0059        \\
$c_{8-link}$ &  .0181  & - .0059 &  -.0107 & .0016        \\
\hline
\end{tabular*}
\end{table*}

We have also constructed an SU(3) action with the same operators. We
fit ``typical'' SU(3) Monte Carlo configurations and a set of
SU(2) instantons. The analog picture of Fig. \ref{fig:newinstsu2}
is shown in Fig.  \ref{fig:newinstsu3}, and a table of the parameter values
is given in Table \ref{tab:eightparsu3}.  We have not done any detailed scaling
tests of this action either. Preliminary values for 
the critical couplings for deconfinement are near
$\beta=2.38$, 2.85 and 3.05 at $N_t=2$, 3, and 4.

These actions are  unwieldy to simulate. 
 The problem is the  high
multiplicity of the perimeter-eight loop (96 distinct paths).
The cost of these actions, compared to the standard Wilson action,
is a ratio of about 225 per link update.  Unfortunately no other loop would
produce even approximately similar scale invariance.  
However, since all the configuration
generation is done on small lattices, generating configurations
is a small part of the computational burden, which we are willing to sacrifice
to the principle that the action should  produce instantons without
scale violations.

\subsection{An ``Algebraic'' Topological Charge Operator}

In this section we first discuss the main problem arising
in connection with the measurement of the topological charge
on the lattice. Then we describe how we deal with it
and also present the details of our algebraic charge operator.

The difficulty is connected to the fact that a topological 
charge can be assigned unambiguously only to continuous
field configurations living in a continuum space-time. In order
to be able to assign a topological charge to a lattice configuration,
we first have to choose an interpolation which maps the lattice
configuration to a continuum configuration. This interpolation
is usually not unique and different interpolations can lead 
to different topological charge assignments. 
If the configuration corresponds to a FP action, Eqn. \ref{STEEP} defines this
interpolation uniquely for the given FP action. (Different FP actions can have
different interpolations and, especially for small instantons, might find
different charge on the same coarse configuration.)
If the configuration does not correspond to a FP action but to a 
renormalized trajectory, 
 one can define the topological charge only statistically  \cite{Wiese}.
This is not surprising as
if one has only incomplete information about a system one usually 
cannot specify the value of an observable but only a probability
distribution of it. In the present case this means that one has  to 
interpolate the lattice fields in all possible ways compatible with the
RG transformation and count 
each interpolation with its Boltzmann weight. In this way one can define a 
probability distribution for the topological
charge on each lattice.

If the lattice configuration is sufficiently smooth this probability
distribution is sharply peaked around a certain value of the
charge and we can assign this charge to the lattice 
configuration. Unfortunately Monte Carlo generated configurations
at typical values of the coupling are not  smooth enough to do this.
It is not surprising then, that it is highly  nontrivial to measure 
the charge on these configurations. For instance the geometric method,
which measures the second Chern number of an interpolated continuum
fiber bundle, gives an integer for any configuration but on these
rough configurations this integer does not have much to do with the topological
charge. We would like to emphasize that this problem is independent 
of the scale invariance or non-invariance of the action we discussed in
Section 2.1. It is present even for FP actions. Rather this problem is the
consequence of the fact that the geometric method was designed to 
work only on smooth configurations where the charge probability 
distribution is sufficiently peaked around an integer and it makes
sense to say that this is {\it the} charge of the given configuration.

In this work we use a  classically and 1-loop
perfect FP action, therefore the inverse
blocking procedure described in Section 2.2 defines a unique 
interpolation to a smoother lattice field with half the lattice spacing 
of the original lattice. The charge of the interpolating field can
be assigned as the charge of  the original rough field. As we shall see later,
in our case after one step of inverse blocking the configuration
becomes smooth enough that a unique integer topological charge
can be defined using either the geometric method
or some improved algebraic operator.

In our earlier work we had measured 
topological charge using the 
 geometric definition\cite{GEO_O3,GEOMETRIC} on the fine configuration.
This definition of the charge 
always gives an integer value and 
has no perturbative corrections \cite{GEO_O3}.

An alternative ``algebraic'' approach to measure the topological charge
 is to introduce a lattice operator $q_a(x)$ which reduces to
the continuum charge operator $F \tilde F$
 for smooth lattice configurations. 
The topological susceptibility measured with this operator  on the lattice 
will receive both a multiplicative and an additive
renormalization constant \cite{Z_BETA}.
The connection between the two methods has been recently discussed
by Rastelli, Rossi, and Vicari \cite{RRV}.
In a recent series of papers a non-perturbative method
was proposed to
evaluate these renormalization constants \cite{APE_TOP}.

We prefer the geometric definition
as it is free of the renormalization constants.
However, measuring the geometric charge is 
a very expensive computational bottleneck.
Our goal was to  find an algebraic definition which reproduces the results of the
geometric charge of the inverse blocked fine configuration on typical
configurations.

We found such an  algebraic charge operator by the following procedure:
We generated a set of ``typical'' configurations and inverse-blocked them.
We measured the geometric charge on the inverse blocked configurations,
as well as the expectation values of several pseudoscalar operators. These operators were
all simple closed loops summed over all possible orientations, with a
sign (plus for even, minus for odd parity transformations
of a certain reference loop) to ensure that the operator was pseudoscalar. 
We then did a least-squares fit to a set of coefficients $c_j$ for a 
set of operators $O_j$, minimizing the chi-squared function
\bee
\chi^2 = \sum_n | \sum_j c_j O_j(n) - Q(n)|^2 .
\ee
where the outer summation is over the different configurations and
$Q(n)$ is the corresponding geometric charge.
This procedure is similar in the spirit of the work of
Ref. \cite{APE_TOP}: we are choosing a linear combination of
operators whose 
multiplicative renormalization factor is effectively
unity, as we are using the geometric
charge as a fiducial.  Were we to change to a very different gauge coupling,
and perhaps to a different gauge group,
this procedure would have to be repeated.

We found a good fit with two operators in two representations
\bee
q_a(x) = \sum_j c_j^1 {\rm Tr}(1-U_j) + c_j^2 ({\rm Tr}(1-U_j))^2
\ee
with operator 1 the path $(x,y,z,-y,-x,t,x,-t,-x,-z)$
 and operator 2 the path $(x,y,z,-x,t,-z,x,-t,-x,-y)$, 
summed over all permutations and reflections.
The coupling coefficients are given in Table \ref{tab:algq}.

\begin{table*}[hbt]
\caption{Couplings of the few-parameter algebraic charge operator
for  SU(2) pure gauge theory.}
\label{tab:algq}
\begin{tabular*}{\textwidth}{@{}l@{\extracolsep{\fill}}lcccccc}
\hline
operator &  $c_1$ & $c_2$  \\
 \hline
1 &  .09030  &  .48846        \\
2 &  -.17863  &  .41270        \\
\hline
\end{tabular*}
\end{table*}

\begin{figure}[!htb]
\begin{center}
\vskip 10mm
\leavevmode
\epsfxsize=90mm
\epsfbox{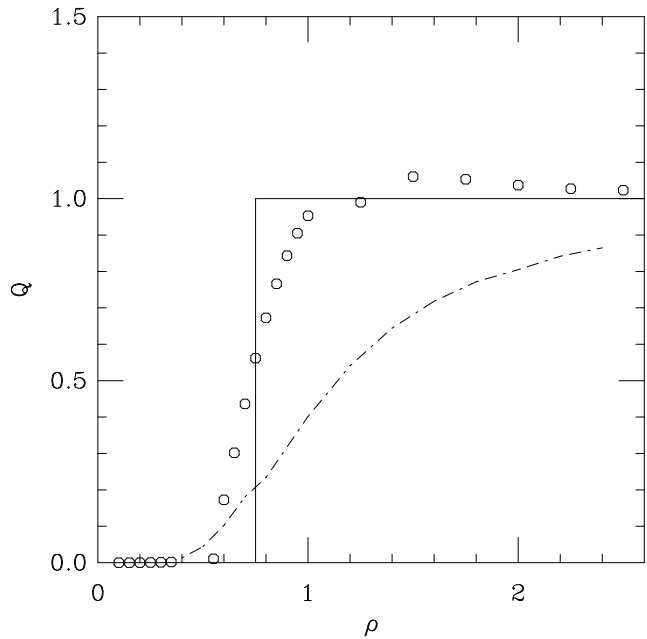}
\vskip 10mm
\end{center}
\caption{ Topological charge vs. instanton radius from our
algebraic charge operator, measured on a set of smooth instanton
configurations. The step function is the geometric
charge on the same configurations. The long-short
dashed line shows the
topological charge measured using the naive definition.}
\label{fig:qvsrho}
\end{figure}

We generated our charge operator 
using typical Monte Carlo configurations. 
Fig. \ref{fig:qvsrho} shows how this operator works on 
a set of very smooth single-instanton configurations.  
 We overlay on this figure the charge
as determined by the geometric method.  
Since we use this charge operator on the inverse blocked configurations where the
instantons are always larger than about twice the lattice spacing, the
improved algebraic charge
deviates from the 
geometric charge by no more than a few per cent.  

We also show on this figure the topological charge for the same set of 
smooth configurations, measured using the ``naive'' 
twisted plaquette $(x$, $y,$ $-x,$ $-y,$ $z,$ $t,$ $-z,$ $-t)$ 
loops.  Over the  range of the plot, it shows a 
marked variation with the instanton radius which can be 
removed by an overall multiplicative
 rescaling $Z(\beta)$ factor only in an  average sense.

We have so far been unsuccessful in constructing an algebraic
topological charge operator which does not involve inverse blocking
as an intermediate step.  

\subsection{Controlled Smoothing}
\label{sec:CS}
A natural way to measure topological properties of the lattice 
vacuum would be to inverse block several times, producing a fine 
interpolation with lattice spacing $(1/2)^n$ of the original coarse
lattice spacing. This program has been carried out for two
dimensional spin models, but the growth 
of the number of variables in a four
dimensional inverse blocking transformation prevents us from making
more than one level of inverse blocking.  Instead, we adopt
the following smoothing procedure:

We first inverse block a coarse lattice to a fine lattice.
During this step, the action
density falls by about a factor of 32,
a factor of 16 because of the increase of the volume and an
additional factor of about 2 because the kernel $\kappa T$
 in Eqn. \ref{STEEP} picks up
about half of the original action. Typical values for the action density in our
simulation are $s_{coarse}=0.9-1.0$  and $s_{fine}=0.02-0.03$ out of 
a maximum value of 2.  
Speaking very naively, this fine  action density corresponds to a
very large ($O(10)$) effective gauge coupling. If we do a
block transformation on a {\it typical}
configuration generated at some  $\beta$ value, the blocked lattice will
correspond to a                                      
$\beta_{eff}\approx \beta - \Delta\beta$ where 
$\Delta\beta=4N_c \beta_0 \ln 2 = 0.257  $, 
($\beta_0 = 11N/(48\pi^2)$, $N_c =2$) 
is given by the continuum $\beta$ function of the model. Starting with a
large $\beta$ configuration the shift  $\Delta\beta$ will have very little
effect on the action density; one expects $s_{blocked} \sim 0.03$.
On the other hand  we know that the fine lattice blocks back to the
original coarse lattice with $s_{coarse} \sim 1.0$. This is due to a
very delicate cancellation and is true only if we block with the same
blocking kernel and define the block links in the same way as for the
inverse block transformation.
The sites of the original coarse lattice correspond to a sublattice of the inverse
blocked fine configuration, in our notation the all-even sublattices, i.e. to
sites with all coordinates even. We have to
perform the block transformation on the fine
lattice with the original blocking kernel and blocked links 
originating from the all-even sublattice, to get back the coarse configuration.
If we do the block transformation with a different blocking
kernel and/or based on a different sublattice, 
these cancellations will not be present, 
and the resulting blocked lattice could be thought of as
corresponding to a large $\beta$ with action density of about 
$s_{block}\sim 0.03$. 

However, to think of the smoothing as merely replacing the coarse lattice
by a fine lattice with a configuration characteristic of very large beta
is incorrect. On
scales larger than the original coarse lattice spacing,
the fine lattice retains all the structure of the coarse lattice.
Only short range fluctuations have been smoothed.
Since the basic assumption behind RGT is that a block transformation does not
change long distance behavior, the physical properties of the blocked lattice
should be the same as the fine lattice and
consequently the same as the original coarse lattice independent of the
choice of block transformation and sublattice. 
The is true for topological objects as blocking does not create or destroy 
instantons larger than some critical size $\rho_c \sim a$. 

With this inverse blocking/blocking procedure we map an original coarse lattice
to a lattice of same size and identical long distance and topological
properties but with greatly reduced action density. The sequence can be repeated
and with each step the action 
density decreases and the resulting configuration becomes
smoother.

We will refer to a blocking-inverse-blocking step as a ``cycle''.
(The first cycle in a sequence is just an inverse blocking step.)
At the end of each cycle, we have a smooth lattice, 
on which we can perform measurements.

In practice, we take our blocking kernel to be 
the same as the inverse blocking kernel
but with $\kappa=\infty$ to reduce fluctuations and build the blocked links
starting from all-odd ((1,1,1,1), etc.) lattice points. In each inverse
blocking step the minimization is done to a precision of less than
1\% of the action of an instanton. With this precision one 
cycle takes about the equivalent (in CPU time) of 15-50 Monte Carlo sweeps 
with our gauge action 
on the coarse lattice. As the cycling proceeds and the configuration
gets smoother the minimization typically becomes faster.
This very precise minimization is probably unnecessary. Prelimenary results
indicate that it might be sufficient to minimize to the accuracy of about 1\% of the
lattice action only. On many times cycled configurations where the action is
dominated by the instantons it does not make much of a difference, but on
rough configurations considerable computer time  can be saved by the relaxed
minimization condition.

\subsection{Properties of the Smoothed Configurations}

As we argued in the previous section, we expect that the many times cycled
configurations have the same physical properties as the original coarse
configuration. In this section we study this problem in more detail.

\begin{figure}[!htb]
\begin{center}
\vskip 10mm
\leavevmode
\epsfxsize=90mm
\epsfbox{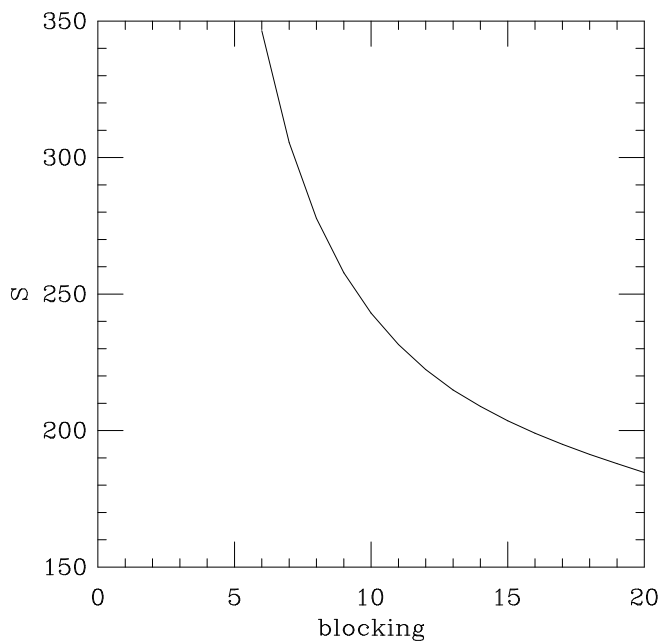}
\vskip 10mm
\end{center}
\caption{ Action vs. cycle number for  a typical 
lattice.}
\label{fig:actioncycle}
\end{figure}

To illustrate the effect of smoothing cycles, 
Fig. \ref{fig:actioncycle} shows the action 
vs. cycle number for a typical $\beta=1.5$ configuration. This particular
configuration has 3 instantons and 2 anti-instanton so the smallest its action
can be is about $S=S_I(A+I)\sim 200$ (assuming the interaction between
instantons is small). The topological charge and the location of the
individual instantons (identified as we describe in Section 2.5) do not
change during the cycling transformation.

\begin{figure}[!htb]
\begin{center}    
\vskip 10mm  
\leavevmode
\epsfxsize=90mm
\epsfbox{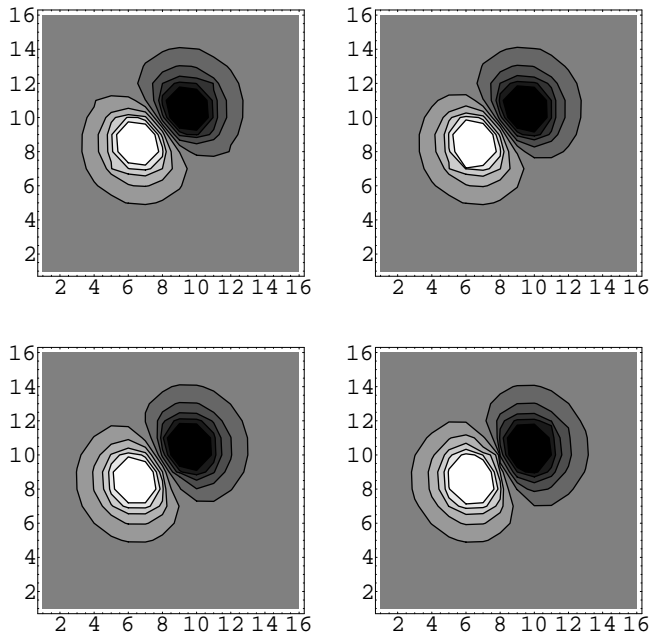}
\vskip 10mm
\end{center}
\caption{The evolution of the charge densitry profile of an ideal I-A pair both
having a radius of $\rho=1.5$ with their centers 2.5 lattice
spacings apart. These snapshots were taken at smoothing step 1,2,6 and
9, always on the inverse blocked lattice.}
\label{fig:pair1} 
\end{figure}

On most configurations we did 9 cycle steps. We measured the topological charge
using the algebraic operator described in Section 2.2 after each cycle.
Occasionally the total charge changed by one signaling the loss of an instanton
or anti-instanton. That is not surprising as instantons that are centered
differently on the lattice have slightly different critical radius
\cite{INSTANTON1}. During a cycling transformation the centers of the instantons
are shifted and small instantons could become unstable. In practice it happens
only for radius $\rho < a$ and it is consistent with the fact that our instanton
identifying algorithm misses instantons with $\rho<\rho_c\sim a$.
Opposite charged pairs are also stable
under the cycling transformation  unless their centers are
closer than about 80\% of the sum of their radii.
We demonstrate this in Fig.\ \ref{fig:pair1} which shows
how the charge profile of an ideal I-A pair evolves through
smoothing.

Next we turn our attention to the static potential. While the cycling
transformation does not change the long distance properties of the system, it
influences the short distance behavior. The more cycling we do, the more
disturbance we will see at short distance. Therefore we expect the string
tension to stay constant but the Coulomb part of the potential can change.

The potential  measurement involves one technical point:
The subtle cancellations on the fine lattice that explain the factor of 32 drop
in the action density after an inverse blocking step create a "staggered"
structure on the fine lattice.   While the
mass gap (exponential falloff of correlators) is the
same on and between  different sublattices, the 
amplitude or overall scale factor in front of the
exponential will be different for correlation functions of fine variables
which are not members of the same sublattices. For example, the 
potential obtained from Wilson loops with length 
even in all directions has a different amplitude than the 
potential obtained from Wilson loops with odd length though the exponential
fall-off for both correlators is the same. 
Mixing different correlators will produce inconsistent, incorrect results. To
avoid the problem one must only
combine correlators of fine variables which occupy a common
sublattice.
In this paper we only present
measurements of the string tension between points of the fine lattice
which belong to the same sublattice,
and we average our measurements over all 16 sublattices.

The results we present in the following were obtained on $16^4$ lattices at
$\beta=1.5$. 

\begin{figure}[!htb]
\begin{center}
\vskip 10mm
\leavevmode
\epsfxsize=90mm
\epsfbox{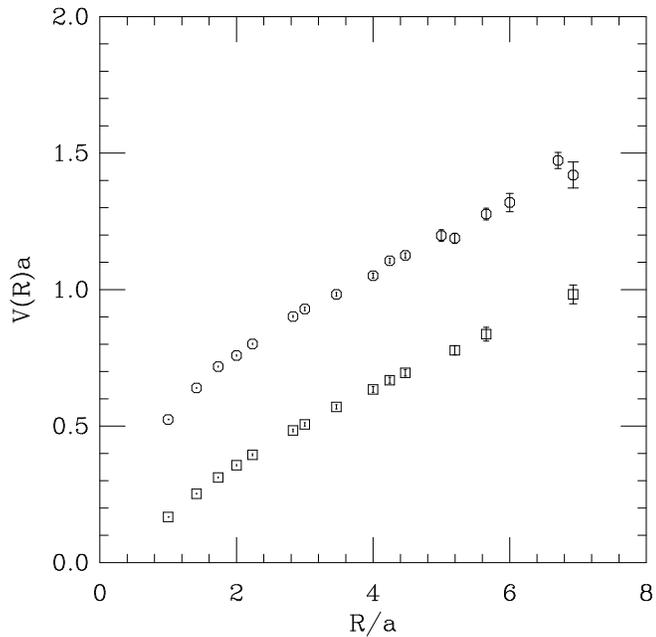}
\vskip 10mm
\end{center}
\caption{ Potential measured on once-inverse-blocked configurations (squares),
compared to standard measurements on the original lattices (octagons),
 for $\beta=1.5$.
}
\label{fig:potib}
\end{figure}

Fig.\ \ref{fig:potib} shows the static potential 
on the original coarse lattice, and then measured on once
inverse blocked lattices. For the latter data set, distances and potentials
 are measured in units of the coarse lattice spacing (units of
two fine lattice spacings).  While the potential on the inverse blocked
lattice is distorted at small lattice spacing, the string tension
is clearly unchanged.  A fit confirms what the eye can 
see--the string tension is unchanged but the short distance Coulomb
part is different.

\begin{figure}[!htb]
\begin{center}
\vskip 10mm
\leavevmode
\epsfxsize=90mm
\epsfbox{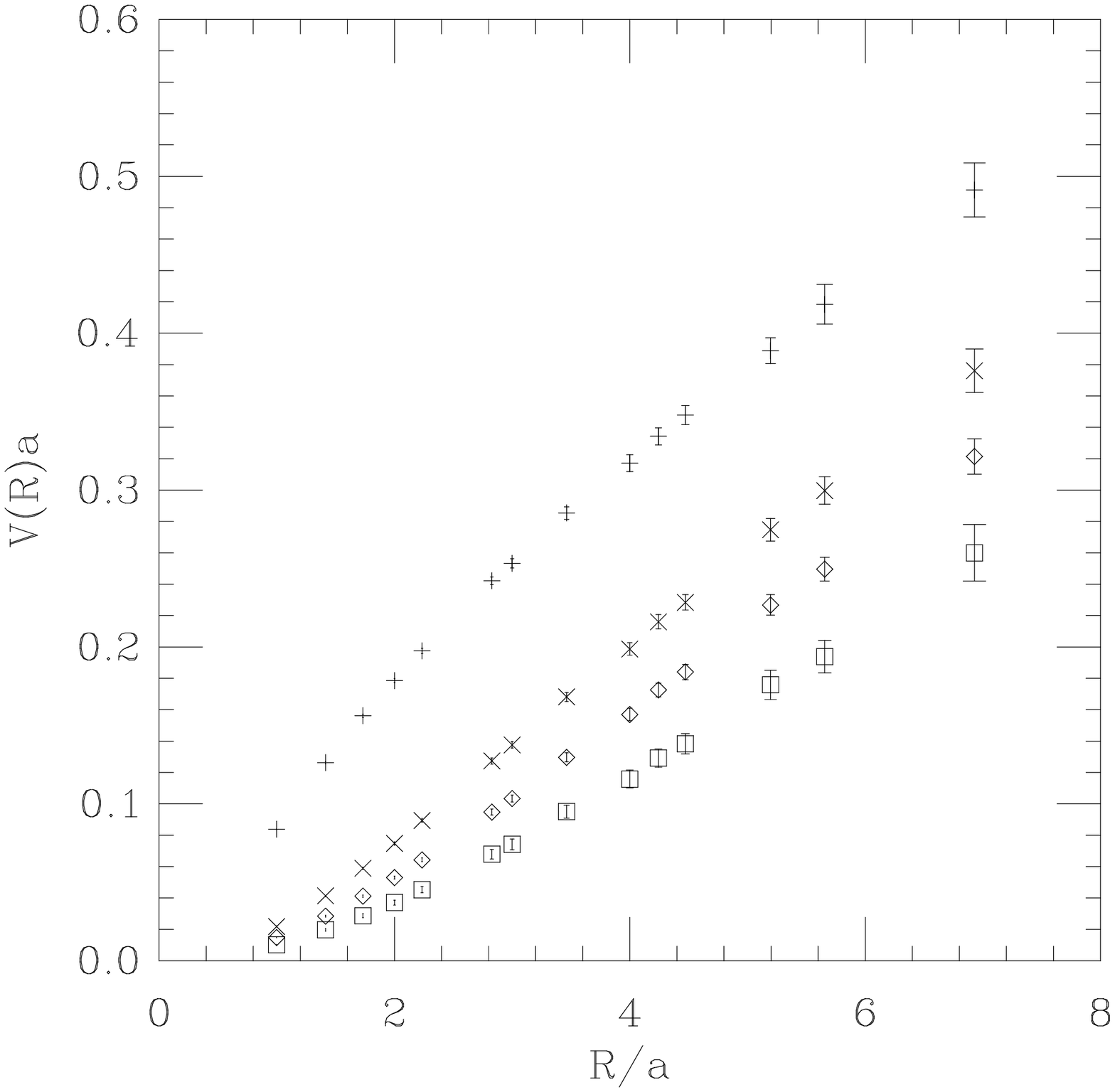}
\vskip 10mm
\end{center}
\caption{ Potential measured on the inverse blocked $16^4$ $\beta=1.5$
configurations after performing different numbers of smoothing cycles
(plusses 1 cycle, crosses 3 cycles, diamonds 5 cycles and squares 9 cycles).}
\label{fig:potcycle}
\end{figure}

On many times cycled configurations we expect the short distance
distortion to increase up to the point that on our relatively 
small lattices the string tension will be distorted as well. 
To test this, we measured how the heavy quark 
potential changes with the number of
smoothing cycles. Figure \ref{fig:potcycle} shows our results obtained on the 
inverse blocked $16^4$ configurations at cycle number 1,3,5 and 9. (We got
exactly the same results by doing the measurement in each cycle on
the blocked $8^4$ lattices.) 

As expected, the behavior of
the potential on length scales of the order of the lattice spacing
changes considerable over the cycles, in particular the Coulomb term is already
quite suppressed on the once inverse blocked configurations.
On the other hand, the string tension, the (constant) slope of the
potential at large distance, does not change at all until 
after about four cycles, when it starts to drop by a few per cent
in each cycle. This is due to the fact that at this point the
effect of smoothing starts to propagate into the distance
scale on which the string tension is measured. We believe that were 
we able to measure the true asymptotic string tension, we would not
notice any change in that.

Recently Feurstein et al.\ \cite{Feurstein} used a similar smoothing
procedure, although they performed the Monte Carlo updates on the fine lattice 
and then blocked and inverse blocked the configurations. They
concluded that one step of blocking and inverse blocking changed the
string tension by 10-25\%, depending on the value of $\beta$.
The reason of the discrepancy between 
our results and theirs might be that they did not distinguish 
the inequivalent sublattices on the inverse blocked configurations.

Thus we conclude that cycling preserves long distance physics
as well as the topological properties of lattice configurations, while
smoothing out short distance fluctuations.

One should contrast this situation with the smoothing method known
as ``cooling,''  \cite{COOL} in which lattice configurations are brought into
equilibrium at very large (or infinite) $\beta$, with the idea that
the short distance fluctuations smooth away first, leaving
long distance structure behind.  While in principle,
because it is a local algorithm,
cooling preserves the long distance structure of a configuration,
in practice the number of cooling steps required to smooth the
configuration enough to see instantons washes away the string tension
\cite{CHUGRANDY}. Also, unless it is done carefully
\cite{VANBAALETC}, cooling distorts the properties of the instantons
as it proceeds, and the net topological charge of a configuration
 can be altered by cooling.

\subsection{Identifying Instantons}

\begin{figure}[!htb]
\begin{center}
\vskip 10mm
\leavevmode
\epsfxsize=90mm
\epsfbox{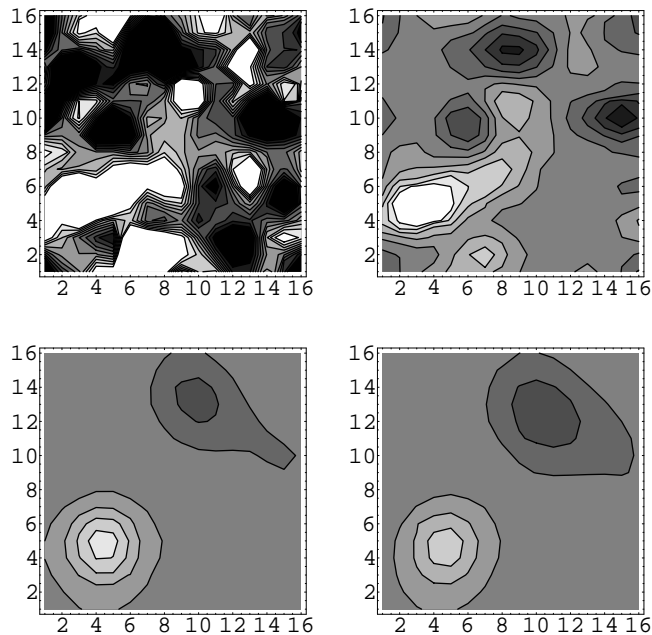}
\vskip 10mm
\end{center}
\caption{The topological charge density in a two-dimensional 
section of a typical configuration at $\beta=1.5$ after 1,2,6 and 
9 smoothing steps.}
\label{fig:sample}
\end{figure}

We identified instantons by looking at the charge density 
measured with our four-parameter algebraic operator on the inverse
blocked fine lattice. We were
looking for peaks of the charge density that had  a density
profile close to that of a single lattice instanton. There is
in principle some arbitrariness in the identification of 
instantons. If an instanton-anti-instanton pair is brought 
closer and closer to one another eventually the positive and 
negative charges cancel and nothing is left. However, this 
happens smoothly and there is no obvious boundary between the
instanton-anti-instanton configuration and the perturbative sector.
Also instantons occurring in real configurations are excited
ones with shapes different from an ideal instanton profile. 
These problems are enhanced by the fact that as we have already
explained, at presently used lattice spacings the configurations
are rather rough. As an illustration we show the charge density 
in a two-dimensional section of one of our once inverse blocked
configurations (Fig. \ref{fig:sample}a). Clearly there is no 
obvious way to identify ``lumps'' in this configuration. Note that
this is a once inverse blocked configuration which is already smooth  
enough that the total topological charge is well defined.
We found that 4-9 cycling steps (depending on the roughness of
the original configuration) was sufficient to smooth the configurations so
that individual instantons could be identified (Fig. \ref{fig:sample}b-d).

To identify instantons we used two different
methods and checked their consistency. The first one was
to look for the largest peaks in the charge density. After 
a peak was found we measured the average charge contained in
spherical shells centered on the peak. Shells of radius up
to 2-3 lattice spacings were used. The charge profile obtained
in this way was compared to that of a set of discretized lattice
instantons with different radii and from this the radius was inferred. 
If the charge profile was not consistent within 5\% 
with the perfect instanton profile the peak was dropped. Also
if a new peak was found closer than the radius of an already 
identified instanton, it was not counted as a separate instanton.
Objects that survived these cuts were identified as instantons and their
charge density was subtracted from the configuration before repeating
the procedure.

According to the second procedure we computed the charge
in spherical shells as above but around each lattice point. 
Then for each site the charge profile around it was fitted
to a (lattice corrected) ideal instanton profile and the
radius was computed from the fit. The fit was done pretending
that there was an instanton center at the given site. Of course
for most of the sites the fit was poor and the obtained radius
very large. However for lattice sites closer and closer to 
instanton centers the fit improved and also the obtained radius
got smaller. It is easy to see that the local minima of
the fitted radius correspond to the actual peaks; if the given 
site is a bit off the peak, the charge density is smaller and 
its average variation between spherical shells of different
size is slower and this results in a larger fitted radius.  
The local {\it minima} of the radius computed in this way identified
instanton centers very well. 
This seemingly awkward procedure was designed to
work on very noisy configurations; for each lattice site we 
collect information from many nearby sites and local fluctuations
could not deceive the algorithm. 

To separate topological objects from
spurious peaks that came mainly from different parts
of the same large instanton we implemented further
consistency checks and cuts. 
If two peaks were closer than the radius of the larger
one, we dropped the one that had a smaller net charge within 
a given radius (typically 2-3 lattice spacings).
If the total charge within a radius of 2-3 lattice spacing
around the peak was less than 50\% of the corresponding 
quantity for an ideal instanton of the same size, the peak 
was discarded.

The apparently arbitrary parameters of our ``cuts'' in both methods
were adjusted by looking at graphs of the charge density 
distribution for about ten configurations at each $\beta$,
identifying the instantons ``by eye'' and making sure that 
the results given by both of the instanton finder algorithms 
were the same. The accuracy of the method can also  be checked 
using the topological susceptibility defined as 
\bee
  \chi_t = \frac{\langle Q^2 \rangle}{V},
\ee
where $Q$ is the total charge in a volume $V$.
The total charge can be measured using the improved charge operator and 
the susceptibility can be computed 
directly. 
On the other hand the charge of each configuration can also be
computed as the number of instantons minus anti-instantons found
by our algorithm. For $\beta=1.5(1.6)$ calculating $Q=I-A$ gives 
$\chi_t=7.5(8)10^{-4}(3.3(3)10^{-4})$ while from the total 
charge measurement  we get $\chi_t=7.0(8)10^{-4}(3.5(3)10^{-4})$. This 
shows that the instanton finder algorithm works well and within 
the error bars it gives the correct susceptibility.

On each configuration we performed 9 cycling steps.
As we have
already mentioned our smoothing algorithm 
left both the location and the radius 
of an ideal lattice instanton unchanged to a very high degree of
accuracy. This was not always true for real 
lattice instantons generated with our Monte Carlo. While their
locations were quite stable from the stage where we could  reliably identify
them, the size of some of the instantons kept changing
slowly throughout the iteration. This change of size was
 less than 20\% during 9 cycling steps for most of the objects, and in most 
cases the instantons grew. To account for this, we 
identified the same lumps from cycling step 6 to step 9 of the iteration
and linearly extrapolated their sizes back to cycling step 1.
Objects whose size changed more than 20\% usually did not have an ideal
instanton profile and often disappeared  after repeated cycling. For
these reasons they are more likely to be quantum fluctuations
and not topological objects, and
 therefore we dropped them from the instanton count.

\section{Results}
\subsection{Parameters of simulations}

We used the static potential
to set the physical scale of
our action. We performed measurements at  $\beta$ values 1.4, 1.5 and 1.6.
On $12^3 \times 16$ lattices
we collected 100-300 measurements of Wilson loops (spaced five update sweeps
apart) per coupling.                                          
The data was fitted
 using the techniques of
Ref. \cite{HELLV} to the static potential $V(r)$ to the form
\bee
   \label{eq:potfit}
V(r) = V_0 + \sigma r - E/r
\ee
In  Table \ref{tab:potential} we show 
the results of the fit together with 
the Sommer \cite{SOMMER} parameter $r_0$
 ($r_0^2 dV(r_0)/dr= -1.65$) 
and we also present the lattice spacing at the different $\beta$ values 
obtained from setting the Sommer parameter to its real-world
value $r_0 = 0.5$ fm. 

Unfortunately at $\beta=1.4$ the configurations were
too rough to measure the string tension and the Sommer
parameter directly on the Monte Carlo generated lattices.
Therefore we measured these parameters on the same set of $16^4$
$\beta=1.4$ once inverse blocked configurations that we
also used for the measurement of the topological charge.
This was done in exactly the same way as was described in 
section \ref{sec:CS}, distinguishing the different
sublattices of the fine configurations. 

Since the inverse blocked lattices correspond to lattice spacing 
one half of the original lattice and the
fluctuations are greatly suppressed, we were able to get 
reliable fits even at relatively large distances and
spatial separations. We measured the potential
this way at $\beta=1.5$ and $1.6$ as well. At $\beta=1.5$ the
results were consistent with the direct calculation. 
At $\beta=1.6$ the $8^4$ lattices showed finite size effects.
  
\begin{table*}[hbt]
\setlength{\tabcolsep}{1.5pc}
\caption{Parameters of the static potential from Wilson loops at 
different values of $\beta$. $\sigma$ is the string tension, 
$r_{min}$ and $r_{max}$ specify the range where the potential was
fitted and $r_0 = 0.5$ fm is the Sommer
parameter. }
\label{tab:potential}
\begin{tabular*}{\textwidth}{@{}l@{\extracolsep{\fill}}lcccccc}
\hline
FP action \\
\hline
$\beta$   & $a^2\sigma$   & $r_0/a$ & $r_{min}$ & $r_{max}$ & $a$ (fm)\\
\hline
 1.4  &  0.216(90) &  2.66(3)  & 1.41  &   4.24 &  0.188(3) \\
 1.5  &  0.122(10) &  3.48(2)  & 1.73  &  10.39 &  0.144(1) \\
 1.6  &  0.078(4)  &  4.30(5)  & 1.73  &  10.39 &  0.116(2)  \\
\hline
\end{tabular*}
\end{table*}

\subsection{Size distribution of instantons and the topological
susceptibility}

The size distribution of instantons in the dilute instanton gas
approximation was calculated in the pioneering paper of 't Hooft \cite{tHooft}
\bee
n(\rho)={C _N\over \rho^5} ({4\pi^2 \over g^2})^{2N} e^{-S_I(g^2)}
\ee
where $n(\rho)d\rho$ is the number of instantons and anti-instanton with
radius between $\rho$ and $\rho+d\rho$, $C_N$ is a regularization
dependent constant and $S_I(g^2)$ is the instanton action with the 
running coupling which at the one loop level reads
\bee
  S_I(g^2(\rho)) =    \frac{8 \pi^2}{g^2(a)}  
                     -\frac{11}{3} N_c \ln(\rho/a).
\ee
For small radius, where this
formula is applicable, $n(\rho)$ increases as a power,
 $n(\rho)\sim \rho^{11N_c/3-5}$. 
For large radius instantons the dilute gas and the perturbative
approximation break down and instanton 
interaction effects will modify $n(\rho)$. One
expects that large instantons are strongly (possibly exponentially)
suppressed in the QCD vacuum but the mechanism of this suppression is
not clear.

We have measured the instanton size distribution on $8^4$ lattices at
$\beta=1.4,1.5$ and 1.6 as described in Section 2.5.
The parameters of our data sets are given in Table \ref{tab:sizedata}. 
The configurations
were separated with 40 Monte Carlo sweeps in each case. In order to compare
different $\beta$ values we present the results in physical units using
for the lattice spacing on the original coarse lattice the values quoted
in Table \ref{tab:potential}.

\begin{table*}[hbt]
\setlength{\tabcolsep}{1.5pc}
\caption{Parameters of the data sets from which topological
 quantities were measured. $L$ is the lattice size (in lattice units)
 and $a$ is the lattice spacing in fm, from string tension measurements.}
\label{tab:sizedata}
\begin{tabular*}{\textwidth}{@{}l@{\extracolsep{\fill}}lcccc}
\hline
FP action \\
\hline
$\beta$   & $L$   & $a$, fm & $L a$, fm & number of configurations \\
\hline
 1.4  &  8 & 0.188  & 1.50 &  84  \\
 1.5  &  8 & 0.144  & 1.15 & 132  \\
 1.6  &  8 & 0.116  & 0.93  & 213  \\
\hline
\end{tabular*}
\end{table*}

\begin{figure}[!htb]
\begin{center}
\vskip 10mm
\leavevmode
\epsfxsize=90mm
\epsfbox{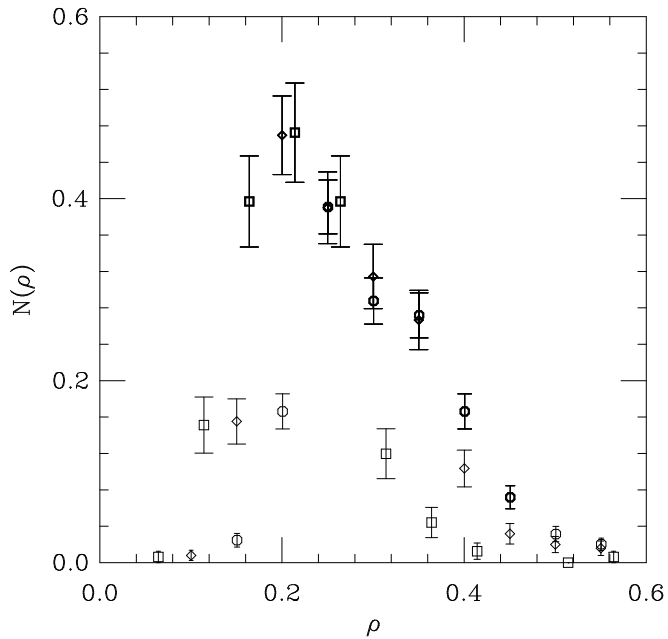}
\vskip 10mm
\end{center}
\caption{ The density distribution of instantons. 
Data at $\beta=1.6$ are given by squares, $\beta=1.5$ diamonds,
and $\beta=1.4$ octagons. The bold data points are ones for which the instanton
radius is large compared to the lattice spacing and small compared to the
simulation volume.}
\label{fig:density}
\end{figure}

In Fig. \ref{fig:density} we present the integrated size distribution
\bee
N(\rho)= {1 \over V} \int_{\rho-\delta \rho/2}^{\rho+\delta \rho/2} n(\rho)d\rho
\ee
where V is the volume in $fm^4$ and 
\bee
\int_{0}^{\infty}n(\rho)d\rho = I+A,
\ee
the total number of instantons and anti-instanton on the lattice.
We chose $\delta \rho=0.05$ fm for all 3 $\beta$ values and chose the bins such
that the third bin corresponded to  $(1.2a,1.2a+0.05)$ fm.
The reason for this choice is to separate regions where our instanton
finding algorithm is reliable ($\rho>a$) from the region where we
probably lose instantons. With the above binning the third bin has all
the physical instantons while the second bin misses some small radius
instantons and the first bin is almost empty.

In addition to the cutoff effects the instanton distribution is
distorted by finite volume effects as well. As large instantons will not
fit into small volumes, $N(\rho)$ will be affected for large $\rho$,
especially for larger $\beta$ values.

If the instanton distribution scales, the results for the 3 different
$\beta$ values should form a universal curve, at least in the region
where finite cutoff and finite volume effects are negligible. In Fig.
\ref{fig:density} we indeed see such a universal curve. The third bins
for $\beta=1.4$ and 1.5 follow this universal curve and that makes us
trust the third bin of the $\beta=1.6$ curve predicting that the density
curve peaks around $\rho\approx 0.2$ fm. For large $\rho$ values the
$\beta=1.6$ curve separates around $\rho\ge 0.3$ fm and we might observe the
$\beta=1.5$ curve separating at around $\rho\ge 0.4$ fm. Combining the three
$\beta$ values we can map the density distribution for 0.15 fm $<\rho<0.5$ fm
which includes most of the interesting region. It would be informative
to go below $\rho=0.15$ fm but for that the calculation has to be repeated
at larger beta values  and on larger lattices. This is not
 an impossible task, but not a workstation project either.

We expected our instanton finding algorithm to lose instantons with
$\rho<a$ and our results are consistent with this expectation. Other
instanton finding algorithms can have different thresholds. For example the
improved cooling method of Ref. \cite{Forcrand} is claimed to keep instantons
above $\rho=2.3a$. This cutoff is about twice  ours. If the simulations
are performed at $a\approx 0.12$ fm, only instantons larger than
$\rho_{min}\approx 0.3$ fm would be counted. This value is well above the peak
of the instanton distribution we found in our simulation. In Ref.
\cite{Forcrand} the distribution peak was found at around $\rho_{peak}\approx
0.45$ fm. In light of the above discussion it is most likely an artificial
peak due to the large $\rho_{cut}$ of the instanton algorithm.

The topological susceptibility can be easily calculated from our data.
We use the usual definition for the susceptibility
\bee
\chi_t = \int d^4x \, \langle q(x)q(0) \rangle
\ee
or
\bee
\chi_t={\langle Q^2 \rangle \over V}
\ee
where Q is the topological charge of the configuration in volume V. 
We do not deal here with the problem of connecting this Euclidean definition
to the Witten-Veneziano formula but we do expect $\chi_t$ to show
scaling.

We quote the result obtained by measuring the total charge
of the configurations using our algebraic operator
$\beta=1.5$.  At $\beta=1.4$ we 
lose small instantons due to the larger lattice spacing, 
and at $\beta=1.6$ we are missing the large
instantons in the simulation. It is possible to combine the three curves
but the result changes very little:
\bee
\chi_t =0.11(1)r_0^4=230(30) \ {\rm MeV}
\ee
This number is consistent with the result we obtained using the FP
action in Ref \cite{INSTANTON2} but it is about 20\%
larger than the quoted susceptibility in
\cite{Forcrand}. Based on the analysis above this discrepancy
 is not surprising. Indeed, if
we keep only instantons with $\rho>0.27$ fm
 we obtain for the susceptibility
$\chi_t=190$ MeV, in complete agreement with the value quoted in
\cite{Forcrand}.

\subsection{Correlations among instantons}

Combining our results for the density distribution at $\beta=1.5$ and
1.6 we find that the instanton density is $n\approx 2.0$ fm$^{-4}$. With
this density and average radius of about $0.2$ fm the instantons are far
from dense. In fact, they occupy only a small fraction, between 3\% and 10\%
of the available space.

\begin{figure}[!htb]
\begin{center}
\vskip 10mm
\leavevmode
\epsfxsize=90mm
\epsfbox{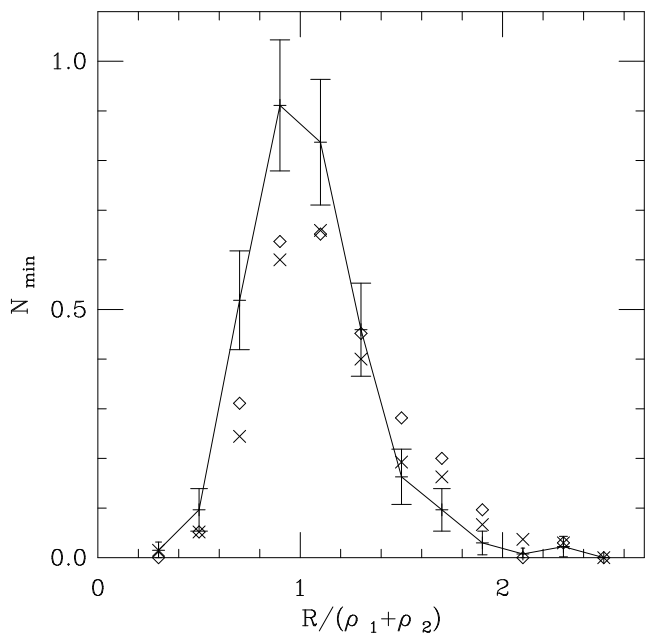}
\vskip 10mm
\end{center}
\caption{ Distribution of nearest objects (solid line). The diamonds correspond 
to the distribution of nearest
like-sign objects, the crosses to nearest unlike-sign
 objects. The error bars for the like and unlike distributions
are suppressed for clarity.
}
\label{fig:mindist}
\end{figure}

Do instantons develop some additional structure preventing them from
growing in size (that would energetically be favorable in the dilute
instanton gas)? We measured the distance between nearest objects to
reveal any correlation. The following analysis was done using only
the $\beta = 1.4$ configurations since these were the only ones
where each lattice contained a significant number of objects (O(10)). 
In Fig. \ref{fig:mindist} we plot the distribution of the
nearest like and unlike objects in addition to the distribution of the
nearest objects  as the function of their distance normalized by their
combined radius, $R/(\rho_1+\rho_2)$.  There is no significant
difference between like and unlike objects. No two objects get closer
than about 80\% of their combined radius indicating a repulsive core
interaction. On the other hand, most objects have a neighbor within 1.2
times of their combined radius. That is significantly closer than the
average distance 0.84 fm inferred from the density.
That could indicate pairing or long-range,
for example linear (polymer like?) structure. To distinguish the two
possibilities we calculated the fraction of instantons belonging to the
largest  cluster on any given lattice. An instanton belongs to a cluster if it
is closer to any member of the cluster than some constant $c$ times 
the combined radius. We found that with $c=1.2$ over
80\% of the instantons belong to the largest cluster indicating 
some type of  long range structure. 

A similar conclusion is reached
by defining the clusters as connected regions of the lattice where 
the charge density is greater than a given number $c$. Measuring $c$
in units of the average (modulus of the) charge density we found that 
for $c=2,3,4$, 95\%, 84\% and 73\% of the instantons belonged to the 
largest cluster. These clustering properties together with the fact that
instantons take up only a small fraction of the total volume show that
the instanton liquid has a very peculiar structure. It is far from 
being close packed; rather, most of the instantons ``touch'' at least one
neighbor and are part of a large cluster.

\subsection{Instantons and confinement}

It is quite remarkable that in our smoothing procedure
after about nine iterations of blocking-inverse blocking, the
action of the resulting lattice configurations is dominated by
instantons, while the long range physics is essentially unchanged.
This means that the large-scale structures that are presumably
responsible for confinement are left intact by our smoothing procedure. 

Since most of the leftover excitations are accounted for by instantons
(at least in terms of action), it is sensible to ask whether the
instantons make a substantial contribution to the heavy quark potential.
It is generally believed that instantons by themselves do not
produce confinement for heavy quarks (for a recent review see 
\cite{Shuryak_long}), however most of the arguments supporting this 
claim use some sort of approximation (e.g.\ dilute instanton gas 
\cite{Callan-DG}, mean field \cite{Diakonov}) that might not be 
justifiable. 

To test this, we measured the contribution of instantons
to the heavy quark potential on the lattice. We generated a 
set of 100 artificial lattice configurations by putting 
together smooth lattice instantons with exactly the same sizes and
locations as we found them in the real Monte Carlo generated
configurations at $\beta=1.5$. Unfortunately we did not have 
any information about the relative orientation of instantons 
in group space so in the artificially built configurations
we put them all aligned. This was done by adding
the vector potentials (the log of the link variables) of 
the smooth instantons in singular gauge. The use of singular
gauge was essential for the fast enough decay of the instanton
vector potentials. In this way we obtained configurations that 
had an action a few percent higher than the sum of the ideal
individual instanton actions. This showed that our ``gluing''
procedure produced smooth configurations. To ensure even more 
smoothness we blocked these $16^4$ configurations to 
$8^4$. 

\begin{figure}[!htb]
\begin{center}
\vskip 10mm
\leavevmode
\epsfxsize=90mm
\epsfbox{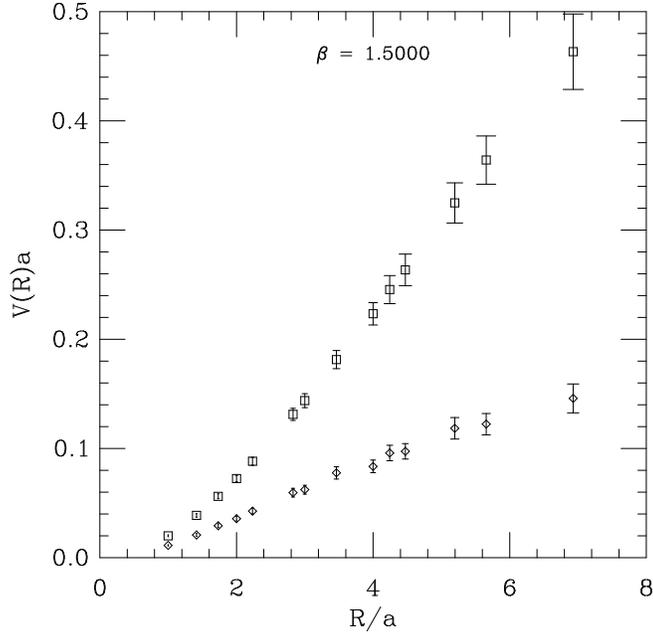}
\vskip 10mm
\end{center}
\caption{The heavy quark potential measured on the artificially 
produced instanton configurations (diamonds) and the potential measured on
the corresponding 9 times smoothed ``real'' configurations (squares),
at $\beta=1.5$.}
\label{fig:inst_pot}
\end{figure}

We measured the heavy quark potential on this ensemble of artificial
configurations and compared it to the same quantity 
measured on the real Monte Carlo generated nine times smoothed
$8^4$ configurations. 
Fig.\ \ref{fig:inst_pot} shows that instantons
make only a very small contribution to the large distance force between
heavy quarks. Our results are in qualitative agreement with the
ones obtained by Diakonov et al.\ \cite{Diakonov_pot} who computed
the heavy quark potential from the instanton liquid. They found that
it goes to a constant at a distance of 1-2 fm. A more quantitative 
comparison of our and their results is not very meaningful since we
computed the potential only up to about 1 fm and also the (fixed)
instanton size that they used was larger than the average size in our
ensemble.

All this, however, shows that instantons by themselves cannot 
explain confinement, and even if they play any role in the confinement
mechanism, it is very subtle.

This might be understood qualitatively as follows. We have already
seen that instantons live at a certain length scale which can be
characterized by their average radius (about 0.2 fm). We expect that
the effect of instantons can be felt only up to this length scale. 
On the other hand confinement, as described by an asymptotic string 
tension, is a genuinely infrared effect and the excitations responsible
for it must be present at all large enough length scales. If 
we perform some gradual coarse graining (e.g.\ by blockings) at 
a certain point the instantons will completely disappear and their 
only remaining trace will be to give some modification to the
effective action for the remaining large scale fluctuations. 
Confinement however is expected to be still present beyond 
this point in the coarse graining procedure. 
 
It follows that the excitations which can be important
for confinement are very large-scale smooth objects that cost 
extremely little in terms of action, but nevertheless  are 
efficient in disordering the system on large distance scales.
These might be for example thick Z(2) vortices \cite{Tomboulis}
or some other semiclassical objects. Another recently proposed 
semiclassical picture of confinement is based on lumps with
topological charge of 1/2 \cite{Arroyo}. Our analysis shows that
on the smoothed configurations the topological structure consists
only of charge (-)1 (anti)instantons and no half charged objects.
It is therefore quite unlikely that charge 1/2 objects play a direct 
role in confinement mechanism.

It would be certainly interesting to know more about the nature
of excitations responsible for confinement. 
Our smoothing technique is well suited to the study of this question because it
suppresses the ultraviolet ``noise'' that dominates the action
on Monte Carlo generated configurations and at the same time it 
does not distort the large scale structure of the lattice 
configurations. We hope to return to this question in a future
publication \cite{Z2}.

\section{Conclusions}
We have described a method for smoothing away the short-distance fluctuations
from equilibrium configurations  of $SU(2)$ gauge field
variables.  The smoothing method preserves topological charge
no matter how much it is used.
It also preserves the long distance structure of
the field configurations, as we show explicitly by measuring the
string tension. The action of the smoothed configurations is dominated
by localized structures--instantons. 
 The density and mean size of the instantons
is quite consistent with the range of expectations from instanton liquid
models. The mean instanton size is quite a bit smaller than previous lattice
simulations of $SU(2)$ gauge theory have reported, and the topological
susceptibility is larger. We believe that this disagreement with
the work of others is due to
the fact that our algorithm for finding topological charge can detect
instantons of size comparable to one lattice spacing, while the other
algorithms have much higher size thresholds.

We find a non-trivial correlation between instantons. They
do not form a dilute gas nor they are closely packed; rather, they appear to clump
in some difficult-to-quantify way.

As far as we can tell, instantons are not directly responsible for
confinement.  Since the smoothing algorithm preserves long distance
physics, we believe that it might be possible to identify 
the field configurations responsible for confinement by examining 
smoothed configurations.
Exactly how that could be done remains an open problem.

Finally, having measured the size distribution and density of instantons
does not tell us whether they are directly responsible for
any observable physics effects.
A major component of instanton phenomenology involves the interactions of
instantons with fermions. A study of those interactions remains an open problem.
We expect that the methods we have described here, especially combined
with an inverse blocking algorithm and fixed-point action for fermions, will
enable us to shed light on these questions.

\section*{Acknowledgements}
We would like to thank P.~Hasenfratz, F.~Niedermayer and A.~Sokal for
useful conversations, and the Colorado High Energy experimental
groups for allowing us to use their work stations.
We would like to thank Matthew Wingate for helping us fit potentials. 
This work was supported by the U.S. Department of 
Energy and by the National Science Foundation.

\newcommand{\PL}[3]{{Phys. Lett.} {\bf #1} {(19#2)} #3}
\newcommand{\PR}[3]{{Phys. Rev.} {\bf #1} {(19#2)}  #3}
\newcommand{\NP}[3]{{Nucl. Phys.} {\bf #1} {(19#2)} #3}
\newcommand{\PRL}[3]{{Phys. Rev. Lett.} {\bf #1} {(19#2)} #3}
\newcommand{\PREPC}[3]{{Phys. Rep.} {\bf #1} {(19#2)}  #3}
\newcommand{\ZPHYS}[3]{{Z. Phys.} {\bf #1} {(19#2)} #3}
\newcommand{\ANN}[3]{{Ann. Phys. (N.Y.)} {\bf #1} {(19#2)} #3}
\newcommand{\HELV}[3]{{Helv. Phys. Acta} {\bf #1} {(19#2)} #3}
\newcommand{\NC}[3]{{Nuovo Cim.} {\bf #1} {(19#2)} #3}
\newcommand{\CMP}[3]{{Comm. Math. Phys.} {\bf #1} {(19#2)} #3}
\newcommand{\REVMP}[3]{{Rev. Mod. Phys.} {\bf #1} {(19#2)} #3}
\newcommand{\ADD}[3]{{\hspace{.1truecm}}{\bf #1} {(19#2)} #3}
\newcommand{\PA}[3] {{Physica} {\bf #1} {(19#2)} #3}
\newcommand{\JE}[3] {{JETP} {\bf #1} {(19#2)} #3}
\newcommand{\FS}[3] {{Nucl. Phys.} {\bf #1}{[FS#2]} {(19#2)} #3}


\end{document}